\documentclass[prl,showpacs,twocolumn,superscriptaddress]{revtex4}
\usepackage{mathrsfs}
\usepackage{amsmath}
\usepackage{amsfonts}
\usepackage{amssymb}
\usepackage{graphicx}
\usepackage{xcolor}
\usepackage{epsfig}

\begin{document}

\title{Coupling superconducting flux qubits at optimal point via dynamic
decoupling from the quantum bus}

\begin{abstract}
We propose a scheme with dc-control of finite bandwidth to implement
two-qubit gate for superconducting flux qubits at the optimal point. We
provide a detailed non-perturbative analysis on the dynamic evolution of the
qubits interacting with a common quantum bus. An effective qubit-qubit
coupling is induced while decoupling the quantum bus with proposed pulse
sequences. The two-qubit gate is insensitive to the initial state of the
quantum bus and applicable to non-perturbative coupling regime which enables
rapid two-qubit operation. This scheme can be scaled up to multi-qubit
coupling.
\end{abstract}

\author{Ying-Dan Wang }
\affiliation{NTT Basic Research Laboratories, NTT Corporation, 3-1,
Morinosato Wakamiya, Atsugi-shi, Kanagawa 243-0198, Japan}
\affiliation{Department of Physics, University of Basel,
Klingelbergstrasse 82, 4056 Basel, Switzerland}
\author{A. Kemp}
\affiliation{NTT Basic Research Laboratories, NTT Corporation, 3-1,
Morinosato Wakamiya, Atsugi-shi, Kanagawa 243-0198, Japan}
\author{K. Semba}
\affiliation{NTT Basic Research Laboratories, NTT Corporation, 3-1, Morinosato Wakamiya,
Atsugi-shi, Kanagawa 243-0198, Japan}
\maketitle


\section{Introduction}

Superconducting Josephson junction (JJ) qubits (for a review, see e.g.~\cite%
{Makhlin2001,YouT2005,Wendin2006,Clarke2008}) provide an arena to
study macroscopic quantum phenomena and act as promising candidates
towards quantum information processing. For the three basic types of
superconducting qubit, namely charge qubit, flux qubit and phase
qubit, single qubit coherent operations with high quality factor
have been demonstrated in many
laboratories~\cite{Nakamura1999,Vion2002,Yu2002,Martinis2002,
Chiorescu2003,Saito2004,Johansson2006,Deppe2008}. However, the best
way to achieve controllable coupling and universal two-qubit gate
are still open questions. A number of experimental
attempts~\cite{Pashkin2003,Yamamoto2003,Berkley2003,Johnson2003,
Izmalkov2004,Xu2005,McDermott2005,Majer2005,Plourde2005,Hime2006,Harris2007,
Ploeg2007,Niskanen2007,Plantenberg2007,Sillanpaa2007,Majer2007} as
well as theoretical proposals have been put
forward~\cite{Makhlin1999,You2002,Averin2003,Wang2004b,Plourde2004,
Rigetti2005,Liu2006,Bertet2006,Niskanen2006,Grajcar2006,Paraoanu2006,
Ashhab2006,Blais2007,Nakano2007} according to the characteristics of
each specific circuit. In this paper, our discussion will be focused
on coupling superconducting flux
qubits~\cite{Mooij1999,Orlando1999,Wilhelm2006}.

The straightforward consideration to realize two-qubit entanglement
is utilizing the fixed inductive coupling between two flux qubits.
With tunable single-qubit energy spacing, this fixed coupling can be
used to demonstrate two-qubit logic gate~\cite{Plantenberg2007}.
However, tunable coupling is required to achieve universal quantum
computing. At early stage, dc-pulse control is widely adopted in the
tunable coupling proposals~\cite{You2005,Plourde2004}. Main
disadvantage for this method is the inefficiency to work at the
degeneracy point which is a low-decoherence sweet spot. At the
optimal point, the natural inductive coupling is off-diagonal in the
diagonal representation of the free Hamiltonian. Hence the coupling
only has second-order effect on the qubit dynamic for the detuned
qubits. Another difficulty related with dc control is the operation
error related to the finite rising-and-falling time of the dc-pulse.
Recently, more attention is paid to coupling schemes with ac-pulse
control~\cite{Rigetti2005,Bertet2006,Liu2006,Niskanen2006,Ashhab2006,Niskanen2007}.
While most of the ac-control coupling schemes can work at the
degeneracy point and no additional circuitry is
needed~\cite{Rigetti2005,Ashhab2006}, some of them require strong
driving~\cite{Rigetti2005} or result in slow
operation~\cite{Ashhab2006}. Meanwhile, unwanted crosstalk is
present due to a always-on coupling. The possible solution to the
above problems is the parametric coupling scheme with a tunable
circuit acting as coupler~\cite{Bertet2006}. A third flux qubit has
been demonstrated as a candidate for this
coupler~\cite{Niskanen2006,Niskanen2007}. However incorporating
additional nonlinear component to the circuit would increase the
complexity of the circuit and might introduce additional noise and
operation errors.

In this paper, we propose a scalable coupling mechanism of flux
qubits with four Josephson junctions in two loops (4JJ-2L). The
coupling is induced by a common quantum bus, such as a LC resonator
or a one-dimensional superconducting transmission line resonator
(TLR). The effective coupling Hamiltonian is diagonal with the free
Hamiltonian of single qubit at the optimal point. With appropriate
dc-control pulse, a dynamic two-qubit quantum gate can be realized
for superconducting flux qubits at the optimal point. The on-and-off
of the coupling can be switched by dc-pulse of finite bandwidth
without introducing additional error. This protocol is based on the
time evolution of a non-perturbative interaction Hamiltonian.
Therefore it is applicable to "ultra strong coupling" regime, where
the coupling strength is comparable to qubit free Hamiltonian.
Contrarily to parametric coupling which requires a strongly
non-linear coupling element, the scheme described in this paper
utilizes a linear element. The strong non-linearity of parametric
coupler required in order to achieve fast enough two-qubit gates
induces strong imperfections of the gates and added the difficulties
related with microwave control~\cite{Niskanen2006}. While the
two-qubit gate based on linear coupler is intrinsically free of
errors if proper DC control is achieved. Thus the linearity of the
coupler considered in this work has not only the advantage to be
insensitive to the state of the coupler, but also offers the
possibility of error-free gates. Due to these advantages, this new
proposal could be a promising alternative in experiments.

This paper is organized as follows. In Sec.~II, we first analyze the
energy spectrum of the 4JJ-2L qubit
configuration~\cite{Mooij1999,Orlando1999}. In Sec.~III, the setup
of our coupling mechanism for this type of qubit is described and
the system Hamiltonian is derived. In Sec.~IV, we present two
different pulse sequences to realize the effective two-qubit
coupling and construct two-qubit logic gates. The characteristics of
this coupling scheme based on experimental consideration are
analyzed in Sec.~V. The discussions of this paper are given in
Sec.~VI.

\section{Flux qubit with tunable qubit gap}

\begin{figure}[bp]
\begin{center}
\includegraphics[bb=214 309 337 542,scale=0.65,clip]{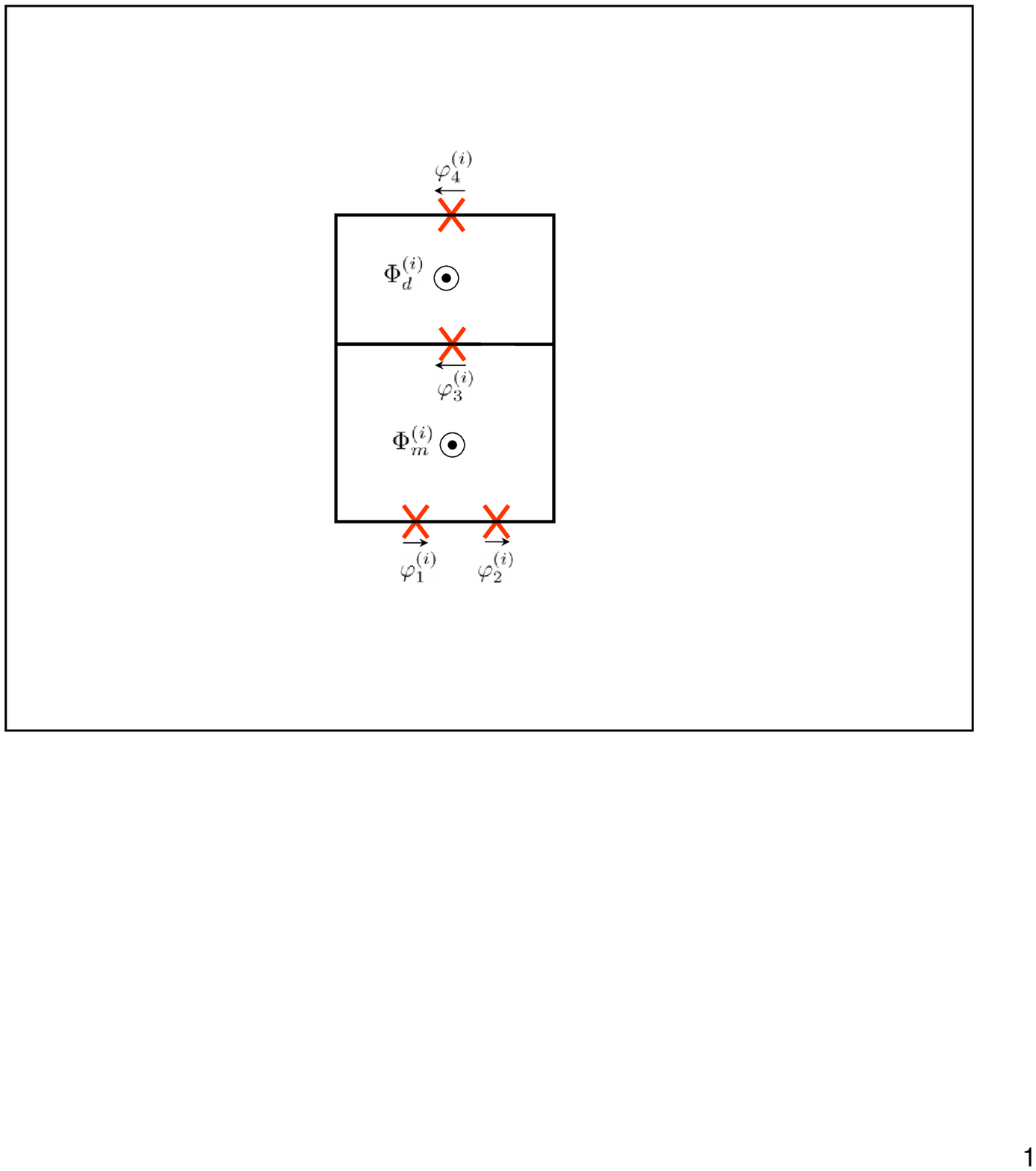}
\end{center}
\caption{(Color online) The schematic of a single qubit with four
Josephson junctions (denoted by cross) connected in two
superconducting loops. The upper loop forms a dc SQUID with two
identical junctions while the lower loop encloses three junctions
similar as the conventional 3-JJ flux qubit. Each loop can be
controlled separately by external magnetic flux
$\Phi^{(i)}_\text{d}$ and $\Phi^{(i)}_\text{m}$ respectively.}
\label{fig:singlequbit}
\end{figure}

A single flux qubit discussed in this paper is shown in
Fig.~(\ref{fig:singlequbit}). Each qubit is composed of four
Josephson junctions in two loops: the main loop (lower loop) and the
dc SQUID loop (upper loop). The main loop encloses three junctions:
two identical junctions with Josephson energy $E_{\text{J}}^{(i)}$
and one shared with the dc SQUID loop with Josephson energy
$\alpha_0^{\left( i\right) }E_{\text{J}}^{\left( i\right) }$ where
$\alpha_0^{(i)}$ is the ratio of the Josephson energy between the
first two junctions and the third one (here and hereafter, the
superscript $(i)$ denotes the variables of the $i$-th qubit). The
main loop forms a flux qubit whose energy eigenstates are the
superpositions of the clockwise and the counterclockwise persistent
current states~\cite{Mooij1999,Orlando1999}. The 4-JJ flux qubit is
different from the conventional design of a flux qubit due to the
additional dc SQUID loop. The third junction of 3-JJ flux qubit is
replaced by a dc SQUID in this 4-JJ design. Therefore the effective
Josephson energy of the third junction can be controlled by the
magnetic flux $\Phi _{\text{d}}^{\left( i\right) }$ threading the dc
SQUID loop. Assuming the two junctions in the dc SQUID loop are
identical, the effective Josephson energy is
$\alpha^{(i)}(\Phi^{(i)}_\text{d})E_\text{J}^{(i)}\equiv
2\alpha_0^{\left( i\right) }\cos \left( \pi \Phi _{\text{d}}^{\left(
i\right) }/\Phi _{0}\right)E_\text{J}^{(i)} $ with $\Phi _{0}$ the
flux quantum. This feature, as we show later, enables the qubit gap
to be tunable. This increases the \emph{in situ} controllability of
the quantum circuit~\cite{Mooij1999,Orlando1999}. The main loop and
the dc SQUID loop of each qubit can be controlled by external
on-site flux bias separately. A high-fidelity two-qubit operation
has been proposed recently for the 4JJ-2L qubit~\cite{Kerman2008}.

As shown in Fig.~(\ref{fig:singlequbit}), the Josephson phase
differences of the four junctions in one qubit are denoted by
$\varphi _{1}^{\left( i\right) }$, $\varphi _{2}^{\left( i\right)
}$, $\varphi _{3}^{\left( i\right)}$ and $\varphi _{4}^{\left(
i\right) }$ respectively. By defining $\tilde{\varphi}_{3}^{\left(
i\right) }\equiv (\varphi _{3}^{\left( i\right) }+\varphi
_{4}^{\left( i\right) })/2$, the total Josephson energy in one qubit
loop is $-U_{0}^{\left( i\right) }=E_{\text{J}}^{\left( i\right)
}\cos \varphi _{1}^{\left( i\right) }+E_{\text{J}}^{\left( i\right)
}\cos \varphi _{2}^{\left( i\right) }+\alpha ^{\left( i\right)
}(\Phi _{\text{d}}^{\left( i\right) })E_{\text{J}}^{\left( i\right)
}\cos \tilde{\varphi}_{3}^{\left( i\right) }$, where we have used
the fluxoid quantization relation in the dc SQUID loop:
\begin{equation}
\varphi _{3}^{\left( i\right) }-\varphi _{4}^{\left( i\right)
}=-2\pi \frac{\Phi _{\text{d}}^{\left( i\right) }}{\Phi _{0}}.
\end{equation}
There are two other fluxoid quantization relations for this circuit:
\begin{eqnarray}
\varphi _{1}^{\left( i\right) }+\varphi _{2}^{\left( i\right)
}+\varphi _{3}^{\left( i\right) } &=&2\pi \frac{\Phi
_\text{m}^{\left( i\right) }}{\Phi _{0}},  \notag \\
\varphi _{1}^{\left( i\right) }+\varphi _{2}^{\left( i\right) }+\varphi
_{4}^{\left( i\right) } &=&2\pi \frac{\Phi _{\text{d}}^{\left( i\right)
}+\Phi _\text{m}^{\left( i\right) }}{\Phi _{0}},  \label{fq}
\end{eqnarray}
where $\Phi _\text{m}^{\left( i\right) }$ the magnetic flux threading the
main qubit loop. Adding up the two equations in (\ref{fq}), we get
\begin{equation}
\varphi _{1}^{\left( i\right) }+\varphi _{2}^{\left( i\right)
}+\tilde{\varphi}_{3}^{\left( i\right) }=2\pi \frac{\Phi
_\text{t}^{\left( i\right) }}{\Phi _{0}},
\end{equation}
where $\Phi _\text{t}^{\left( i\right) }\equiv\Phi _\text{m}^{\left(
i\right) }+\Phi _{\text{d}}^{\left( i\right) }/2$ is the total magnetic flux
threading the qubit loop. Then the total Josephson energy of the four
junctions in the loop is
\begin{eqnarray}
-U_{0}^{(i)}&=& \alpha ^{\left( i\right) }(\Phi _{\text{d}}^{\left(
i\right) })E_{\text{J}}^{(i)}\cos \left( 2\pi \frac{\Phi
_\text{t}^{\left( i\right) }}{\Phi _{0}}-\left( \varphi _{1}^{\left(
i\right) }+\varphi _{2}^{\left(
i\right) }\right) \right)  \notag \\
&&+E_{\text{J}}^{(i)}\cos \varphi _{1}^{\left( i\right)
}+E_{\text{J}}^{(i)}\cos \varphi _{2}^{\left( i\right) }. \label{U0}
\end{eqnarray}
It takes the same form as that of the 3-JJ flux
qubit~\cite{Mooij1999,Orlando1999} except that the ratio $\alpha
^{\left( i\right) }$ is tunable. If the total magnetic flux $\Phi
_\text{t}^{\left( i\right) }$ is close to half a flux quantum $\Phi
_{0}/2$ and $\alpha ^{(i)}>0.5$, the function $U_{0}\left( \varphi
_{1}^{\left( i\right) },\varphi _{2}^{\left( i\right) }\right) $
represents a landscape with periodic double-well potentials.

With external flux bias, one can set the operation point in one double-well
potential. The classical stable states of this potential correspond to the
clockwise and the counter-clockwise persistent current states. By changing
the ratio $\alpha^{\left( i\right) }$ between the Josephson energy of the
third junction (through the dc SQUID), the height of the tunneling barrier
(hence the tunneling rate) between the two minima of each double-well is
tunable. When $\alpha^{\left( i\right) }$ is set in appropriate range,
coherent tunneling between the two wells of the potential is enabled while
the tunneling between different potentials is highly suppressed.

\begin{figure}[bp]
\begin{center}
\includegraphics[bb=184 278 406 506, scale=0.65,clip]{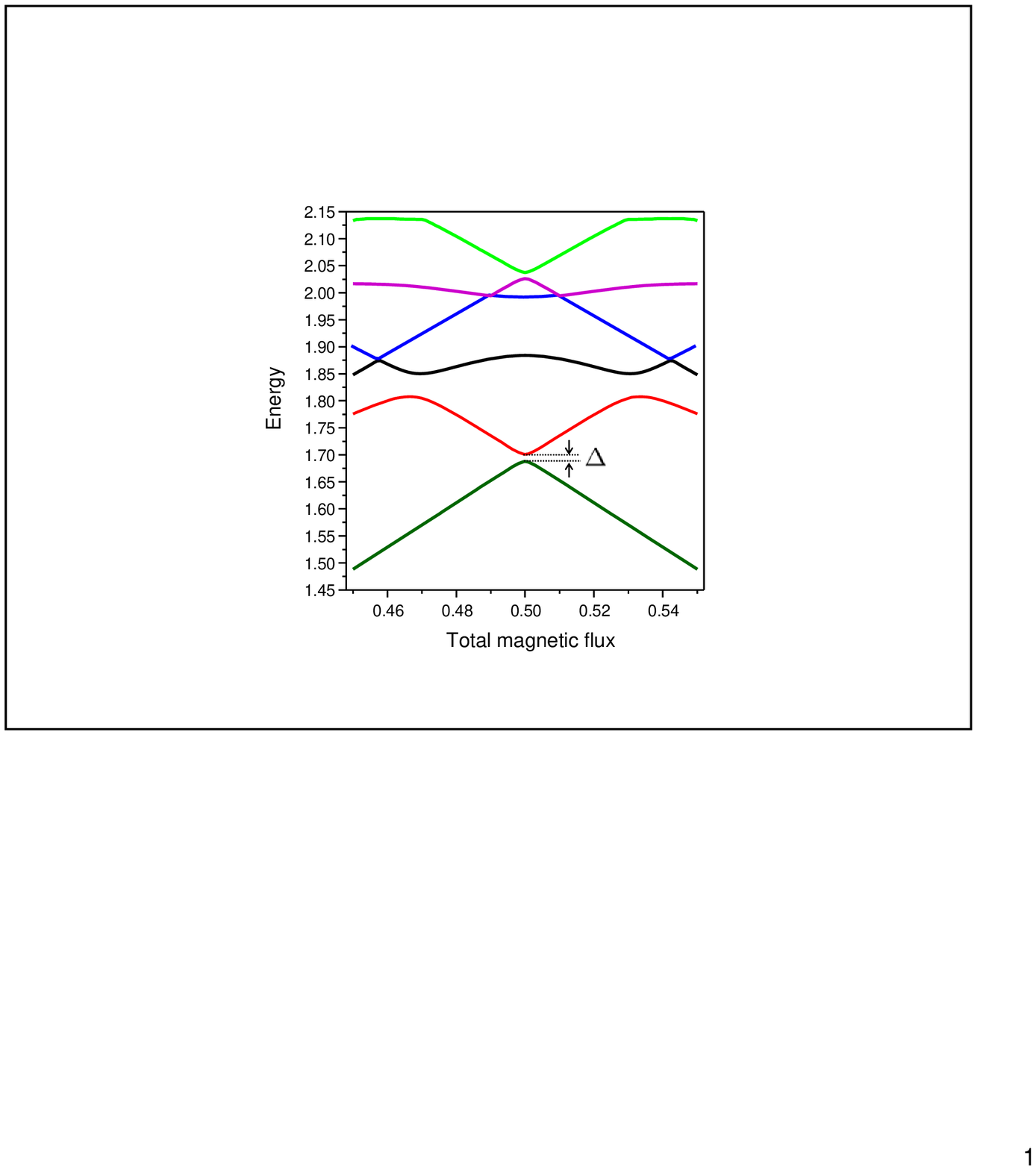}
\end{center}
\caption{(Color online) The energy spectrum of the lowest six energy
levels of the superconducting loop with respect to the total
magnetic flux $\Phi _\text{t}^{(i)}$. The energy is in the unit of
$E_{\text{J}}$ while the magnetic flux is in the unit of $\Phi
_{0}$. We take $E_{\text{J}}^{(i)}/E_\text{C}^{(i)}=35$ and
$\protect\alpha ^{(i)}=0.8$.} \label{fig:spectrum}
\end{figure}

\begin{figure}[tp]
\begin{center}
\includegraphics[bb=85 293 483 560, scale=0.6,clip]{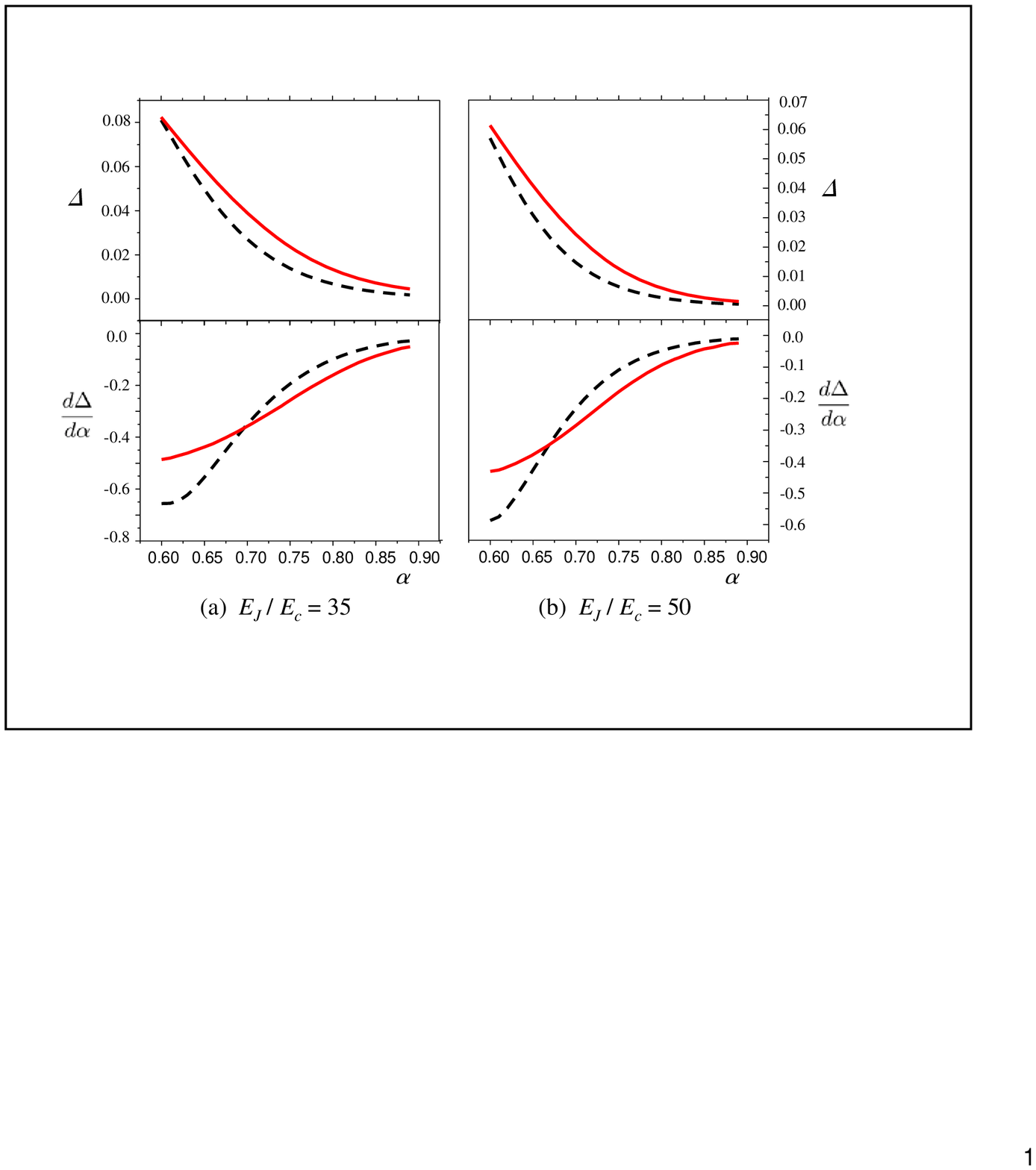}
\end{center}
\caption{(Color online) The energy gap of single qubit $\Delta
^{(i)}$ and its derivative $d\Delta ^{(i)}/d\protect\alpha ^{\left(
i\right) }$ as a function of $\protect\alpha ^{\left( i\right) }$
for (a) $E_{\text{J} }^{(i)}/E_{\text{C}}^{(i)}=35$ and (b)
$E_{\text{J}}^{(i)}/E_{\text{C}}^{(i)}=50$ (for simplicity, the
superscript $(i)$ is omitted in the figure). The solid line (red) is
obtained from the exact diagonalization of the original 4-JJ qubit
Hamiltonian while the dashed line (black) is obtained from the
analytical solution of the tight-binding model with WKB
approximation which breaks down at low barrier regime. The energy is
in the unit of $E_\text{J}$.} \label{fig:gap}
\end{figure}

Taking into account the electric energy stored in the four
capacitors, we can get the full Hamiltonian of this system. The
energy spectrum of the circuit with $\alpha ^{\left( i\right) }=0.8$
and $E_\text{J}^{(i)}/E_\text{C}^{(i)}=35$
($E_\text{C}^{(i)}=e^2/2C$ denotes the Coulomb energy of the first
(second) junction of the $i$-th qubit and $C$ is the junction
capacitance) is shown in Fig.~\ref{fig:spectrum} as a function of
the rescaled total magnetic flux $f^{\left( i\right) }=\Phi
_\text{t}^{\left( i\right) }/\Phi _{0}$.
In the vicinity of $\Phi _\text{t}^{\left( i\right) }=\Phi _{0}/2$,
the lowest two energy levels are far away from other energy levels
and form a two-level subspace which can be used as a flux qubit. The
eigenstates of the flux qubit are superpositions of the clockwise
and the counter-clockwise persistent current states. The 4-JJ flux
qubit works the same as its 3-JJ prototype except that the barrier
height of the double-well potential is tunable \textit{in situ}. In
the two-level subspace, the free Hamiltonian for the $i$-th qubit is
written as
\begin{equation}
H^{\left( i\right) }=\frac{\varepsilon ^{\left( i\right) }(\Phi
_\text{t}^{\left( i\right) })}{2}\sigma _{{z}}^{\left( i\right)
}+\frac{\Delta ^{\left( i\right) }(\Phi _{\text{d}}^{\left( i\right)
})}{2}\sigma _{x}^{\left( i\right) }  \label{Hi}
\end{equation}
where $\varepsilon ^{\left( i\right) }$ is the energy spacing of the two
classical current states
\begin{equation}
\varepsilon ^{\left( i\right) }(\Phi _\text{t}^{\left( i\right) })\approx 2
I_{p}^{\left( i\right) }\left( \Phi _\text{t}^{\left( i\right) }-\frac{\Phi
_{0}}{2}\right)
\end{equation}
and $\Delta ^{\left( i\right) }$ is the energy gap between the two states at
the degeneracy point $\Phi _\text{t}^{(i)}=\Phi _{0}/2$,
\begin{equation}
\Delta ^{\left( i\right) }(\Phi _{\text{d}}^{\left( i\right) })\equiv \Delta
^{\left( i\right) }(\alpha ^{\left( i\right) }=2\alpha _{0}^{\left( i\right)
}\cos (\pi \frac{\Phi _{\text{d}}^{\left( i\right) }}{\Phi _{0}})).
\end{equation}
According to the tight-binding model, $\Delta^{(i)}$ can be
evaluated through WKB approximation~\cite{Orlando1999} as
$\Delta^{(i)}\approx( \omega_a/2\pi)\exp(-[4\alpha(1+2\alpha)
E_\text{J}^{(i)}/E_\text{C}^{(i)}]
^{1/2}(\sin\phi^{*(i)}-\phi^{*(i)}/2\alpha))$ where $\omega_a$ is
the attempt frequency of escape in the potential well and
$\cos\phi^{*(i)}=0.5\alpha^{(i)}$ (the Planck constant $\hbar$ is
set to be $1$). In Fig.~\ref{fig:gap}, the energy gap $\Delta
^{(i)}$ and its derivative $d\Delta ^{(i)}/d\alpha ^{\left( i\right)
}$ are shown as a function of $\alpha ^{\left( i\right) }$. The
results are obtained from numerical calculation and analytical
derivation based on WKB approximation.

\section{The Coupled System}

\begin{figure}[tp]
\begin{center}
\includegraphics[bb=154 290 436 555,scale=0.65,clip]{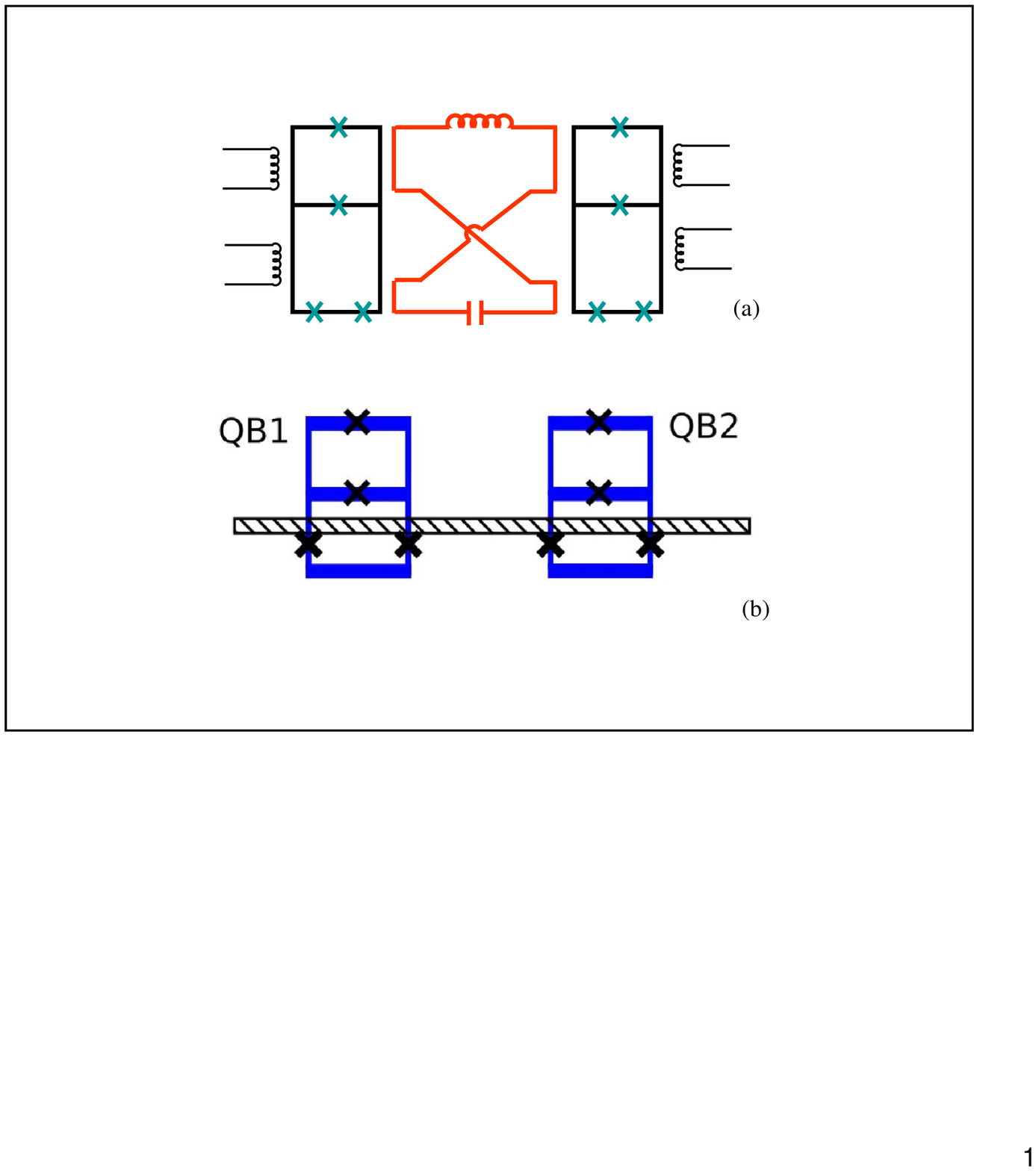}
\end{center}
\caption{(Color online) The circuit design examples to implement the
required coupling. Two 4JJ-2L flux qubits are coupled with each other
through the inductive coupling with a resonator as data bus: (a) a twisted
LC resonator and (b) a 1D superconducting transmission line resonator in a
separate layer. The current of the data bus induces magnetic fluxes both in
the upper loop and in the lower loop of each qubit. The directions of two
magnetic fluxes are opposite.}
\label{fig:setup}
\end{figure}

A schematic to illustrate our coupling mechanism is shown in
Fig.~\ref{fig:setup} with two different types of data bus, i.e., LC
resonator and 1D TLR. For simplicity, we first concentrate on
coupling two qubits. The problem of scale-up will be discussed
later. As we described in the previous section, each qubit is
composed of four Josephson junctions in two loops: the main loop
(the lower loop) and the dc SQUID loop (the upper one). The main
loop and the dc SQUID loop of each qubit can be controlled by
external on-site flux bias independently. The two qubits are placed
in sufficient distance so that the direct coupling can be
effectively neglected~\cite{Niskanen2007}. The two qubits are both
coupled with a common data bus such as a twisted LC resonator or 1D
on-the-top TLR via mutual inductance.

Due to the mutual inductance with the resonator, the magnetic fluxes include
the contribution both from the external applied flux and the resonator,
i.e., $\Phi _\text{m}^{\left( i\right) }=\Phi _{\text{m,e}}^{\left( i\right)
}+\Phi _{\text{m,b}}^{\left( i\right) }$ and $\Phi _{\text{d}}^{\left(
i\right) } =\Phi _{\text{d,e}}^{\left( i\right) }+\Phi _{\text{d,b}}^{\left(
i\right) }$, where the subscript $e (b)$ indicates the contribution from the
external magnetic flux (the quantum data bus) respectively. Then the total
magnetic flux $\Phi _\text{t}^{\left( i\right) }$ reads
\begin{equation}
\Phi _\text{t}^{\left( i\right) }=\left( \Phi _{\text{m,e}}^{\left(
i\right) }+\frac{\Phi _{\text{d,e}}^{\left( i\right) }}{2}\right)
+\left( \Phi _{\text{m,b}}^{\left( i\right) }+\frac{\Phi
_{\text{d,b}}^{\left( i\right) }}{2}\right).  \label{ph1}
\end{equation}

The coupling between a single qubit and the data bus includes two
parts: the coupling of the qubit with the dc SQUID loop via mutual
inductance $M_{\text{d}}^{\left( i\right) }$ and the coupling of the
qubit with the main loop via $M_\text{m}^{\left( i\right) }$. The
magnetic flux induced in the dc SQUID loop and the qubit main loop
are
\begin{eqnarray}
\Phi _{\text{d,b}}^{\left( i\right) } &=&M_{\text{d}}^{\left( i\right) }I
\notag \\
\Phi _{\text{m,b}}^{\left( i\right) } &=&M_\text{m}^{\left( i\right) }I
\label{f1}
\end{eqnarray}
respectively and $I$ is the current in the resonator. For our
purpose, the two magnetic fluxes satisfy
\begin{equation}
\Phi _{\text{d,b}}^{\left( i\right) }=-2\Phi _{\text{m,b}}^{\left( i\right)
}.  \label{cond1}
\end{equation}
This can be implemented by designing the mutual inductance
\begin{equation}
M_{\text{d}}^{\left( i\right) }=-2M_\text{m}^{\left( i\right) },  \label{mco}
\end{equation}
The minus in (\ref{cond1}) is due to the special layout of the data
bus so that the directions of the magnetic flux induced by the
quantum bus in the upper loop and the lower loop are opposite.
Inserting Eq.~(\ref{cond1}) into Eq.~(\ref{ph1}), we find the total
flux $\Phi _\text{t}^{\left( i\right) }$ is contributed only by the
external applied magnetic flux as $\Phi _\text{t}^{\left( i\right)
}=\Phi _{\text{q,e}}^{\left( i\right) }+\Phi _{\text{d,e}}^{\left(
i\right) }/2$. Since the $\sigma _{z}$ component of the qubit is
coupled with $\Phi _\text{t}^{\left( i\right) }$, the resonator
contributes a pure $\sigma _{x}$ coupling with no $\sigma _{z}$
component. Therefore the qubit can always be biased at the optimal
point $\Phi_\text{t}^{(i)}=\Phi_0/2 $.

For the quantized mode of the resonator,
\begin{equation}
I=\sqrt{\frac{\omega }{2L}}\left( a+a^{\dag }\right) .
\end{equation}
where $\omega =(LC)^{-1/2}$ is the plasma frequency of resonator,
$L$ ($C$) the lumped or distributed inductance (capacitance) of the
resonator and $a^{\dag }$ ($a$) the plasmon creation (annihilation)
operator. With these denotations,
\begin{equation}
\Phi _{\text{d,b}}^{\left( i\right) }=f_{\text{d}}^{\left( i\right) }\left(
a+a^{\dag }\right)
\end{equation}
where
\begin{equation}
f_{\text{d}}^{\left( i\right) }\equiv M_{\text{d}}^{\left( i\right)
}\sqrt{ \frac{\omega }{2L}}.
\end{equation}
Usually the mutual inductance of the resonator and the qubit loop is
about several pH to several tens of pH. For example, if we take
$M_{\text{d}}^{\left( i\right) }=10$ pH, $\omega =1$ GHz and $L=100$
pH, $f_{\text{d}}^{\left( i\right) }/\Phi_{0}\approx 5.6\times
10^{-4}\ll 1$. This means the magnetic flux contributed from the
resonator is much smaller than that from the external applied
magnetic field. To the first order, the energy gap of a single qubit
is modified by the resonator as
\begin{equation}
\Delta \left( \alpha ^{\left( i\right) }\right) \approx \Delta \left( \alpha
_{e}^{\left( i\right) }\right) -\left. \frac{d\Delta \left( \alpha ^{\left(
i\right) }\right) }{d\alpha ^{\left( i\right) }}\right\vert _{\alpha
^{\left( i\right) }=\alpha _{e}^{\left( i\right) }}\delta \alpha ^{\left(
i\right) }(a+a^{\dag})  \label{linear}
\end{equation}
with
\begin{eqnarray}
\alpha _{e}^{\left( i\right) } &=&2\alpha _{0}^{\left( i\right) }\cos \left(
\pi \frac{\Phi _{\text{d,e}}^{\left( i\right) }}{\Phi _{0}}\right)  \notag \\
\delta \alpha ^{\left( i\right) } &=&2\alpha _{0}^{\left( i\right) }\pi \sin
\left( \pi \frac{\Phi _{\text{d,e}}^{\left( i\right) }}{\Phi _{0}}\right)
\frac{f_{\text{d}}^{\left( i\right) }}{\Phi _{0}}
\end{eqnarray}
The Hamiltonian for a single qubit linearly interacting with the data bus
reads,
\begin{eqnarray}
H^{\left( i\right) }&=&\frac{\varepsilon ^{\left( i\right) }\left(
\Phi _{m,e}^{\left( i\right) }+\Phi _{\text{d,e}}^{\left( i\right)
}/2\right)}{2} \sigma _{z}^{\left( i\right) }+\frac{\Delta ^{\left(
i\right) }\left( \Phi _{\text{d,e}}^{\left( i\right) }\right)
}{2}\sigma _{x}^{\left( i\right) }
\notag \\
&& + g^{\left( i\right) }\left( \Phi _{\text{d,e}}^{\left( i\right) }\right)
\sigma _{x}^{\left( i\right) }\left( a+a^{\dag }\right)
\end{eqnarray}
with
\begin{equation}
\Delta ^{\left( i\right) }\left( \Phi _{\text{d,e}}^{\left( i\right)
}\right)\equiv \Delta \left( \alpha _{e}^{\left( i\right) }\right),
\end{equation}
and the coupling coefficient
\begin{equation}
g^{\left( i\right)}\left( \Phi _{\text{d,e}}^{\left( i\right)
}\right)\equiv\kappa^{\left( i\right)}\left( \Phi
_{\text{d,e}}^{\left( i\right) }\right)\sqrt{\omega},\label{gi}
\end{equation}
with
\begin{equation}
\kappa^{\left( i\right)}\left( \Phi _{\text{d,e}}^{\left( i\right)
}\right) = -\left. \frac{d\Delta \left( \alpha ^{\left( i\right)
}\right) }{d\alpha ^{\left( i\right) }}\right\vert _{\alpha ^{\left(
i\right) }=\alpha _{e}^{\left( i\right) }} \delta \alpha ^{\left(
i\right) }.
\end{equation}
Note that magnitude of the
coupling $g^{(i)}$ increases with the mutual inductance
$M_{\text{d}^{(i)}}$. If $\Phi _{m,e}^{\left( i\right) }+\Phi
_{\text{d,e}}^{\left( i\right) }/2=(n+0.5)\Phi _{0}$ (where $n=0,
\pm 1, \pm 2$ is an arbitrary integer), qubit is biased at the
degeneracy point and the system Hamiltonian is written as
\begin{equation}
H=\omega a^{\dag }a+\sum_{i=1,2}\left( \frac{\Delta ^{\left( i\right) }(\Phi
_{\text{d,e}}^{\left( i\right)} )}{2}\sigma _{x}^{\left( i\right)
}+g^{\left( i\right) }(\Phi _{\text{d,e}}^{\left( i\right) })\sigma
_{x}^{\left( i\right) }\left( a+a^{\dag }\right) \right).  \label{hami}
\end{equation}
By tuning the external magnetic flux $\Phi _{\text{d,e}}^{\left(
i\right) }$ in the dc SQUID loop to be$\ n\Phi _{0}$, $g^{\left(
i\right) }(\Phi _{\text{d,e}}^{\left( i\right) })=0$, the qubit is
decoupled from the resonator in the first order. The qubits act
independently and single-qubit operation can be implemented by
biasing $\Phi_{\text{q,e}}^{\left( i\right) }$ together with
microwave pulse.

In the above discussion, the condition Eq.~(\ref{mco}) is assumed. However
it might not be precisely satisfied in practical case. Suppose there is a
small deviation in the fabrication process that $M_{\text{d}}^{\left(
i\right) }=-2(1+\delta )M_{\text{m}}^{(i)}$ (where $\delta \ll 1$), the
total magnetic flux $\Phi _{\text{t}}^{\left( i\right) }$ includes a small
contribution from the resonator,
\begin{equation}
\Phi _{\text{t}}^{\left( i\right) }=\Phi _{\text{q,e}}^{\left(
i\right) }+\frac{\Phi _{\text{d,e}}^{\left( i\right) }}{2}-\delta
M_{\text{m}}^{\left( i\right) }I.
\end{equation}
This adds a term to the Hamiltonian Eq.~(\ref{hami}): $g^{\prime
\left( i\right) }\sigma _{z}^{\left( i\right) }\left( a+a^{\dag
}\right) $ with $g^{\prime \left( i\right) }=-\delta
M_{\text{m}}^{\left( i\right) }I_{p}\sqrt{{\omega }/{2L}}$. However
since the qubit is far-detuned (e.g. according to the parameters
used in Sec. V, $\Delta\approx 15.28$ GHz and $\omega \approx 1$
GHz), this last term is a fast-rotating one and has negligible
contribution. In the following, we adopt Eq.~(\ref{hami}) as the
effective system Hamiltonian.

\section{The strategy to achieve effective two-qubit interaction}
In this section, we discuss about how to achieve the two-qubit
coupling in this composite system. The qubits only interact with
each other indirectly through a common quantum bus. In the
dispersive limit, the operation time of two-qubit logic gate is
limited by the small ratio $g/{\delta\omega}$ where $g$ is the
qubit-bus coupling and $\delta\omega$ the qubit-resonator detuning.
In this case, the resonator is only virtually excited. In this paper
we rely on another stradegy that the quantum bus carries real
excitations. The effective two-qubit coupling is achieved by one or
a series of specific unitary evolutions of the resonator-qubit
composite system. Similar method has been discussed in the context
of quantum computing with thermal
ion-trap~\cite{Milburn1999,Sorensen1999,Sorensen2000,Wangxg2001} and
Josephson charge qubit~\cite{Wang2004b}. The feature of this
coupling is the insensitivity to the quantum state of the resonator.
In ion-trap quantum computing, it is known as S{\o}renson-M{\o}lmer
gate and has been experimentally
demonstrated~\cite{Sackett2000,Leibfried2003,Haljan2005,Home2006}.
However, the original S{\o}renson-M{\o}lmer relies on virtual
excitations of the vibrational modes, whereas here the quantum bus
carries real excitations.

If a dc-pulse is applied to $\Phi _{\text{d,e}}^{(i)}$ to shift it
from $n\Phi _{0}$, the time evolution of the composite system is
driven by the Hamiltonian Eq.~(\ref{hami}). The operators included
in the interaction Hamiltonian $(a+a^{\dag })\sigma _{x}^{(i)}$ and
the free Hamiltonian ($a+a^{\dag }$, $\sigma _{x}^{(i)}$) may be
enlarged by their commutator into a closed Lie algebra of finite
dimension. Thus the exact solution of the time evolution can be
decomposed into a product over exponentials of the
generators~\cite{Wei1963}. In the interaction picture,
\begin{equation}
H_{I}=\sum_{i}g^{\left( i\right) }\left( a^{\dag }e^{i\omega t}+ae^{-i\omega
t}\right) \sigma _{x}^{\left( i\right) }.
\end{equation}
The corresponding closed Lie algebra is $\left\{ a\sigma _{x}^{\left(
i\right) },a^{\dag }\sigma _{x}^{\left( i\right) },\sigma _{x}^{\left(
1\right) }\sigma _{x}^{\left( 2\right) },1\right\} $. The time evolution
operator as the product of their exponentials can be written as
\begin{eqnarray}
U_{\text{I}}(t) &=&e^{-iD\left( t\right) }e^{-iA(t)\sigma _{x1}\sigma
_{x2}}\cdot   \notag \\
&&\left( \prod\limits_{i=1,2}e^{-iB_{i}(t)a\sigma _{x}^{\left(
i\right) }}\right)\left(\prod\limits_{i=1,2} e^{-iB_{i}^{\ast
}(t)a^{\dag }\sigma _{x}^{\left( i\right) }}\right) , \label{u1}
\end{eqnarray}
where
\begin{equation}
\left\{
\begin{array}{c}
B_{i}(t)=\frac{g^{\left( i\right) }}{-i\omega }\left( e^{-i\omega
t}-1\right) , \\
A\left( t\right) =\frac{2g^{\left( 1\right) }g^{\left( 2\right) }}{\omega }
\left( \frac{1}{i\omega }\left( e^{i\omega t}-1\right) -t\right) , \\
D\left( t\right) =\frac{(g^{\left( 1\right)
})^{2}+(g^{(2)})^{2}}{\omega } \left( \frac{1}{i\omega }\left(
e^{i\omega t}-1\right) -t\right) .
\end{array} \right.
\label{p1}
\end{equation}
In the following discussion, we neglect the universal phase factor $D\left(
t\right) $. If the last factor of Eq.~(\ref{u1}) can be effectively
canceled,
\begin{equation}
U_{\text{I}}\left( t\right) \equiv \exp [-iD\left( t\right) -iA(t)\sigma
_{x1}\sigma _{x2}],  \label{u2}
\end{equation}
which represents the time evolution which is effectively governed by
Hamiltonian $\sim \sigma _{x1}\sigma _{x2}$. This can be done in two
different ways as described below:

\begin{figure}[bp]
\begin{center}
\includegraphics[bb=14 650 368 814,scale=0.6,clip]{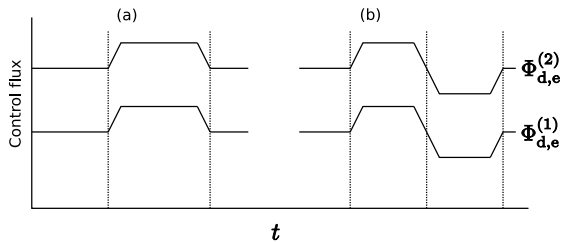}
\end{center}
\caption{(Color online) Schematics of the pulse sequence to realize
two-qubit gate operation.} \label{fig:pulse}
\end{figure}

\subsection{Single pulse operation}

By controlling the pulse length, a two-qubit gate is realized with a
single dc-pulse which shifts $\Phi_{\text{d,e}}^{(i)}$ from
$n\Phi_0$ (Fig.~\ref{fig:pulse} (a)). While the whole time evolution
Eq.~(\ref{u1}) is non-periodic, Eq.~(\ref{p1}) shows that $B_{i}(t)$
is a periodic function of time and it vanishes at $T_{n}={2n\pi
}/{\omega }$. At these times, the time evolution operator in the
interaction picture reduces to
\begin{equation}
U_{\text{I}}\left( T_{n}\right) =\exp \left( i4n\pi \kappa^{\left(
1\right) }\kappa^{\left( 2\right) }\sigma _{x1}\sigma _{x2}\right) .
\end{equation}
This is equivalent to a system of two coupled qubits with an interaction
Hamiltonian $\propto \sigma _{x1}\sigma _{x2}$.

The minimum time to realize a rotation $U_{xx}(\theta )=\exp \left( i\theta
\sigma _{x1}\sigma _{x2}\right) $ is
\begin{equation}
T_{\min }\equiv T_{m_{0}}=2m_{0}\pi /\omega
\end{equation}
with
\begin{equation}
m_{0}=\left[ n_{0}\right] =\left[ \frac{\theta\omega}{4\pi
\kappa^{\left( 1\right) }\kappa{\left( 2\right) }}\right] ,
\end{equation}
where $\left[ ...\right] $ represents the integer part of a number.
Note that we can not achieve a two-qubit rotation precisely unless
$n_{0}$ happens to be an integer, so that $n_{0}=m_{0}$. The error
of one two-qubit gate is of the order of $4\pi \kappa^{\left(
1\right) }\kappa^{\left( 2\right) }/\omega $ (about $1\%$ using
practical parameters). This operation error can be avoided using a
double-pulse method discussed below. It is notable that increasing
the frequency of the resonator $\omega $ cannot achieve a faster
two-qubit gate (due to the $\sqrt{\omega }$ dependence of $g^{\left(
i\right) }$), but it can reduce the error of the two qubit gate.

\subsection{Double pulse operation}

Alternatively a two-qubit logic operation can be constructed with two
successive operations as shown in Fig.~\ref{fig:pulse}(b).

Initially $\Phi _{\text{d,e}}^{(i)}$ is biased at $n\Phi _{0}$. The
first dc-pulse shifts it to a certain $\Phi _{\text{d,e}}^{(i)}\neq
n\Phi _{0}$ for a duration ${t}/{2}$. The evolution operator (in the
interaction picture) is $U_{\text{I}}\left( \frac{t}{2}\right) $.
After time $t/{2}$, one reverses the direction of the magnetic flux
in the dc SQUID loop so that $\Phi _{\text{d,e}}^{\left( i\right) }$
is changed into $-\Phi _{\text{d,e}}^{\left( i\right) }$ and
$g^{\prime \left( i\right) }=-g^{\left( i\right) }$. The system is
then driven by a new Hamiltonian $H^{\prime }=H\left( -g^{\left(
i\right) }\right) $ for another ${t}/{2}$. The time evolution during
the second pulse is
\begin{eqnarray}
U_{I}^{^{\prime }}\left( t\right)  &=&e^{-iD^{^{\prime }}\left( t\right)
}e^{-iA^{^{\prime }}(t)\sigma _{x1}\sigma _{x2}}\cdot   \notag \\
&&\left(\prod\limits_{i=1,2}e^{-iB_{i}^{\ast ^{\prime }}(t)a^{\dag
}\sigma _{xi}}\right)\left(\prod\limits_{i=1,2}e^{-iB_{i}^{^{\prime
}}(t)a\sigma _{xi}}\right) \nonumber \\
&& \label{u4}
\end{eqnarray}
with
\begin{equation}
\left\{
\begin{array}{c}
B_{i}^{^{\prime }}(t)=-B_{i}(t) \\
A^{^{\prime }}\left( t\right) =\frac{2g_{1}g_{2}}{\omega }\left(
\frac{1}{-i\omega }\left( e^{-i\omega t}-1\right) -t\right)  \\
D^{^{\prime }}\left( t\right) =\frac{g_{1}^{2}+g_{2}^{2}}{\omega }\left(
\frac{1}{-i\omega }\left( e^{-i\omega t}-1\right) -t\right)
\end{array}
\right.
\end{equation}
Note that the two terms in the brackets of Eq.~(\ref{u4}) are
permuted comparing with Eq.~(\ref{u1}), so that the expressions
$A'(t)$, $D'(t)$ are different from $A(t)$, $D(t)$ in
Eq.~(\ref{p1}).

The dynamics for the above two consecutive steps is
\begin{eqnarray}
U_{\text{tot}}\left( t\right)  &=&U_{I}^{\prime }\left( \frac{t}{2}\right)
U_{I}\left( \frac{t}{2}\right)  \\
&=&\exp [-iM(t)\sigma _{x1}\sigma _{x2}]\exp [-iN\left( t\right) ],
\end{eqnarray}
where
\begin{eqnarray}
M(t) &=&\frac{2g^{\left( 1\right) }g^{\left( 2\right) }}{\omega }\left(
\frac{2}{\omega }\sin \frac{\omega t}{2}-t\right) ,  \notag \\
N(t) &=&\frac{(g^{\left( 1\right) })^{2}+(g^{(2)})^{2}}{\omega
}\left( \frac{2}{\omega }\sin \frac{\omega t}{2}-t\right) .
\end{eqnarray}
Therefore, the total time evolution is equivalent to the time evolution
governed by two-qubit interaction $\sim \sigma _{x}^{\left( 1\right) }\sigma
_{x}^{\left( 2\right) }$ together with a universal phase factor.

The time $T$ to realize a rotation $U_{xx}(\theta )$ in this way satisfies
the nonlinear equation
\begin{equation}
\frac{\omega T}{2}-\sin \frac{\omega T}{2}=\frac{\theta
\omega}{4\kappa^{\left( 1\right) }\kappa^{\left( 2\right) }}.
\label{t1}
\end{equation}
In the case of $\omega \gg g^{\left( i\right) }$, the solution is written as
\begin{equation}
T\approx \frac{\theta}{2\kappa^{\left( 1\right) }\kappa^{\left(
2\right) }}.
\end{equation}

The two-qubit operation time is estimated using experimental parameters.
Assuming $M_{\text{d}}=10$ pH and $L=100$ pH, one gets $g^{\left( i\right)
}\approx 36.02\text{ MHz}$. As a cost of the low fluctuation related to the
dc SQUID loop, the coupling strength associated with the dc SQUID loop is
weaker than that with the main loop. For example, to realize a $U_{xx}\left(
\frac{\pi }{2}\right) $, the operation time is about $204$ ns. It is smaller
than the qubit coherence time at the optimal point. The operation time is
proportional to $L/M_{\text{d}}^{2}$. Increasing the mutual inductance
between the dc SQUID and the resonator reduces the operation time. It is
worth to point out that the ratio $g/\omega $ is not required to be small.
Therefore there is no fundamental limit on the operation time except the
realizable coupling strength.

As discussed in Sec. III, arbitrary single qubit gate can be
performed after switching off the qubit-bus interaction. Any
non-trivial two-qubit gate can be built up with this xx coupling
plus single qubit gates. For example, the C-phase gate can be
constructed as (up to a global phase factor) \cite{Makhlin2001}
\begin{equation}
R\left( \theta \right) \equiv U_{z}^{\left( 1\right) }\left(
-\frac{\theta }{2}\right) U_{z}^{\left( 2\right) }\left(
-\frac{\theta }{2}\right) \exp \left( i\theta
\tilde{\sigma}_{z}^{\left( 1\right) }\tilde{\sigma}_{z}^{\left(
2\right) }/4\right)
\end{equation}
with $U_{z}^{\left( i\right) }\left( \theta \right) \equiv \exp
\left( i\theta \tilde{\sigma}_{z}^{\left( i\right) }/2\right) =\exp
\left( i\theta \sigma _{x}^{\left( i\right) }/2\right) $. Here we
change representation so that $\tilde{\sigma}_{z}=\sigma _{x}$ and
$\tilde{\sigma}_{x}=-\sigma _{z}$. And the CNOT gate can be readily
constructed with C-phase gate as
\begin{equation}
\text{CNOT}=H^{\left( 2\right) }R\left( \pi \right) H^{\left( 2\right) },
\end{equation}
where $H^{(i)}$ denote the Hadamard transformation on the $i$-th
qubit as $H^{\left( i\right) }\equiv \exp \left( -i\pi \left( \sigma
_{x}^{\left( i\right) }+\sigma _{z}^{\left( i\right) }\right)
/2\sqrt{2}\right) $ (up to a phase factor).

With arbitrary single-qubit rotation and any non-trivial two-qubit rotation,
universal quantum computing can be realized according to quantum network
theorem~\cite{Barenco1995}.

\section{The features of this coupling protocol}

In the previous section, we have presented the way to realize two-qubit
coupling and logic gate with our proposed setup. In this section, the
features of this coupling protocol are analyzed with emphasis on the
experimental implementation. The qubit-qubit effective coupling commutes
with the free Hamiltonian of the single qubit. This feature enables many
practical advantages:

(1) The main idea to implement a two-qubit operation from the exact
evolution operator Eq.~(\ref{u1}) is to cancel the part related with
the degree of freedom of the resonator, so that the final operation
Eq.~(\ref{u2}) represents a qubit-qubit operation without
entanglement with resonator mode. Therefore the resonator mode does
not transfer population with the qubit although the resonator mode
mediates the qubit-qubit interaction. As a result, this two-qubit
logic gate is insensitive to the initial state of the
resonator~\cite{Sorensen1999}. This feature is important for the
experiment performed at finite temperature because the equilibrium
state of the resonator is a mixed state. For example, there is
$16\%$ population at the excited state for a $1$ GHz resonator at
$30$ mK.

(2) As we mentioned, our coupling protocol works at the low-decoherence
optimal point where the qubit is robust to flux fluctuation and has long
decoherence time. This is in contrast to other coupling protocols with
dc-pulse control~\cite{You2005,Plourde2004}. During the two-qubit operation,
the control parameter is not the total magnetic flux but rather a component
in the dc SQUID loop. Therefore, the qubit can be biased at the optimal
point $\Phi _\text{t}^{\left( i\right) }=(n+1/2)\Phi _{0}$ during two-qubit
operation.

While the dc SQUID adds a second control to the circuit, it
introduces extra decoherence. The fluctuation of the flux threading
the dc SQUID loop results in the fluctuation of the energy splitting
$\Delta $ and introduces decoherence to the qubit dynamics. Suppose
the magnetic flux are perturbed by the same amount of fluctuation as
$\Phi _{\text{t}}^{(i)}\rightarrow \Phi _{\text{t}}^{(i)}+\delta
\Phi $, $\Phi _{\text{d}}^{(i)}\rightarrow \Phi
_{\text{d}}^{(i)}+\delta \Phi $. Therefore the first-order effect of
the fluctuation of magnetic flux in the main loop and the sub-loop
are $\delta E_{\text{t}}^{(i)}\equiv c_{\text{t}}^{(i)}\delta \Phi $
and $\delta E_{\text{d}}^{(i)}\equiv c_{\text{d}}^{(i)}\delta \Phi $
respectively, with $c_{\text{t}}^{(i)}\equiv \left\vert \partial
E^{(i)}/\partial \Phi _{\text{t}}^{(i)}\right\vert $ and
$c_{\text{d}}^{(i)}\equiv \left\vert \partial E^{(i)}/\partial \Phi
_{\text{d}}^{(i)}\right\vert $, where $E^{(i)}$ is the energy level
spacing of the qubit, $E^{(i)}=\sqrt{\varepsilon ^{2}\left( \Phi
_{\text{t}}^{(i)}\right) +\Delta ^{2}\left( \Phi
_{\text{d}}^{(i)}\right) }$. If a qubit~\cite{Chiorescu2003} with
$E_{\text{J}}/\hbar =259$ GHz, $E_{\text{J}}/E_{c}=35$ and $2\alpha
_{0}=0.8$ is biased at $\Phi _{\text{t}}=\Phi _{0}/2$ and $2\alpha
_{0}\cos \left( \pi \Phi _{\text{d}}/\Phi _{0}\right) \approx 0.65$,
we get
\begin{equation}
c_{\text{d}}^{(i)}=163\text{ GHz}/\Phi _{0}.
\end{equation}
However, if qubit is not biased at the optimal point but close to the
optimal point, e.g. $\varepsilon /E=0.5$,
\begin{equation}
c_{\text{t}}^{(i)}=1100\text{ GHz}/\Phi _{0}
\end{equation}
where we assume $I_{p}=500$ nA. The influence of the fluctuation on
the total magnetic flux is one order of magnitude larger than that
on the dc SQUID loop. This suggests that although the dc SQUID loop
introduces additional fluctuation to the system, the decoherence
coming from flux fluctuation in dc SQUID is much less than that
caused by shifting away from the degeneracy point $\Phi
_{\text{t}}^{\left( i\right) }=\Phi _{0}/2$.

(3) A scalable qubit-qubit coupling scheme should allow the coupling
to be switched on-and-off (i.e. tunable over several orders of
magnitude). Otherwise, additional compensation pulse is needed to
correct the error in single-qubit operation. In our coupling
protocol, as shown in Eq.~(\ref{gi}), the external magnetic flux in
the dc SQUID loop can be used to switch off the coupling by setting
$\Phi _{\text{d,e}}^{\left( i\right) }=2n\pi \Phi _{0}$. When the
qubit is decoupled from the data bus, single qubit operation can be
controlled by $\Phi _{\text{q,e}}^{\left( i\right) }$ independently.

Our protocol does not require to change the amplitude of a dc pulse
instantaneously. Finite rising and falling times of the controlling
dc pulse will not induce additional error to the two qubit coupling.
This is essentially due to the qubit-resonator interaction commutes
with the free Hamiltonian of the qubit at the optimal point. In the
previous discussion, we assumed a constant $g^{\left( i\right) }$
for simplicity. In the experiments, the modulation of the magnetic
flux always needs finite rising time, i.e., $g^{\left( i\right)
}=g^{\left( i\right) }\left( t\right) $. As long as $g^{\left(
i\right) }$ is a slow-varying (comparing with $ e^{-i\Omega t}$)
function of time $t$, the above discussion still holds except that
the length of the pulse, i.e. $T$ should satisfy
\begin{equation}
B_i\left( T\right) \equiv e^{-i\omega T}g^{(i)}\left( T\right)
-g^{(i)}\left( 0\right) =0  \label{c2}
\end{equation}
instead of $T=2n\pi /\omega $. The magnitude of the effective two-qubit
interaction, i.e., $A(t)$ in Eq.~(\ref{u1}) is modified as
\begin{eqnarray}
A\left( T\right) &=& \int_{0}^{T}\frac{dt}{\omega }\{e^{i\omega t}(
g^{(1)}\left( t\right) g^{(2)}\left( 0\right)+g^{(1)}\left( 0\right)
g^{(2)}\left( t\right))  \notag \\
&& -2g^{(1)}\left( t\right) g^{(2)}\left( t\right) \} .
\end{eqnarray}
To realize a certain xx rotation $U=\exp \left( i\theta \sigma
_{x}^{\left( 1\right) }\sigma _{x}^{\left( 2\right) }\right) $ is to
apply a pulse satisfy $B_i\left( T\right) =0$ and $A\left( T\right)
=\theta $ simultaneously. It is notable that the two conditions are
only related to the integral over the whole pulse and thus robust to
operation error. This conclusion is also applicable to the
double-pulse method.

(4) The evaluation is applicable to "ultra-strong coupling" regime
where the coupling strength is even comparable to the free
Hamiltonian frequency as long as the approximation~(\ref{linear}) is
valid. Hence in principle, the two-qubit operation can be made as
fast as single qubit operation.

\section{Discussion and Conclusions}
\begin{figure}[bp]
\begin{center}
\includegraphics[bb=105 309 418 542,scale=0.65,clip]{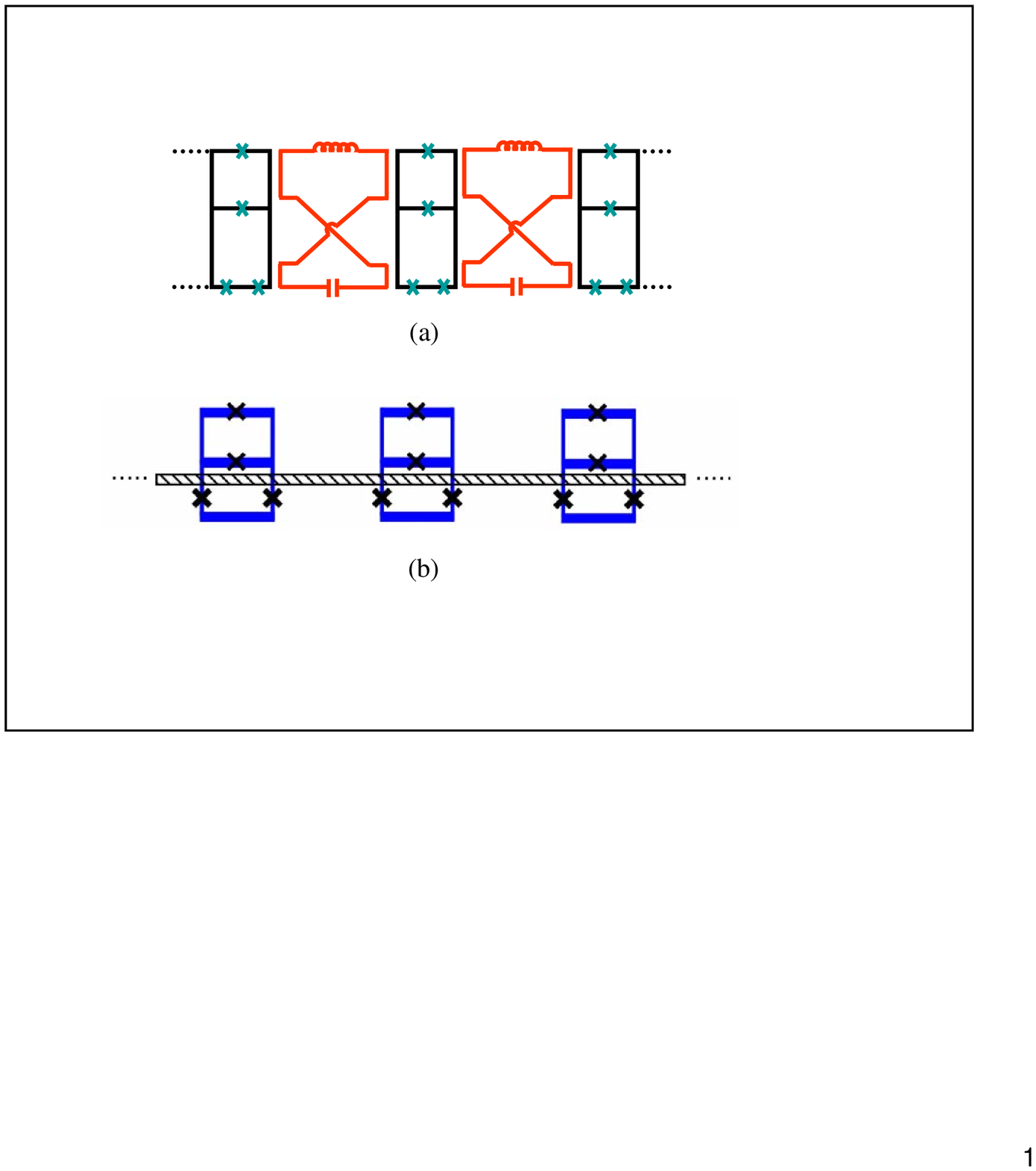}
\end{center}
\caption{(Color online) Schematics to scale up the coupling system.
(a) Each qubit is coupled with the nearest neighbors by twisted LC
resonators. (b) All qubits are interacting with a common TLR
resonator on top of the qubits array.} \label{fig:multiqubit}
\end{figure}
We illustrate two possible ways to scale up the two-qubit system. In
Fig.~\ref{fig:multiqubit}~(a), a nearest-neighbor coupled qubits are
sketched. They form a transverse Ising chain which can be used to
implement quantum state
transfer~\cite{Bose2003,Song2005,Lyakhov2005} and quantum
information storage~\cite{Wang2005}. It is possible to extend this
configuration to 2D Ising model. Fig.~\ref{fig:multiqubit}~(b) shows
an example to realize selective coupling between multiple qubits by
a single quantum bus (such as a transmission line resonator).

For nominally same parameters, there is natural spread of the
junctions critical currents. This coupling mechanism is robust to
the difference of $\alpha _{0}^{\left( 1\right) }$ and $\alpha
_{0}^{\left( 2\right) }$ because the free Hamiltonian commutes with
the interaction Hamiltonian. As such, in the sample fabrication
process, the requirements on homogeneity and reproducibility can be
relaxed and meet with current production technology. The additional
on-site control lines require only one more layer.

The qubit-resonator interaction commutes with the qubit free
Hamiltonian. This feature enables quantum non-demolition (QND)
measurement on superconducting qubit biased at the optimal
point~\cite{Blais2004}. This QND measurement is realizable even in
the ultra-strong coupling limit~\cite{Lupascu2007}.

\section*{Acknowledgement}

The authors are very grateful to S. Saito and H. Nakano for their
suggestions on the experimental realizations of this proposal. YDW
also thank C. P. Sun, Yu-xi Liu and Fei Xue for fruitful discussions
and J. Q. You for his assist on numerical calculation. This work was
partially supported by the JSPS-KAKENHI No. 18201018 and
MEXT-KAKENHI No. 18001002. YDW was also supported by EC IST-FET
project EuroSQUIP.


\end{document}